\RequirePackage[hyphens]{url}
\documentclass[referee]{raa}
\usepackage{amsmath,amssymb,natbib}
\usepackage{graphicx,paralist,comment}
\usepackage[a4paper,margin=1in]{geometry}

\let\Re\relax
\DeclareMathOperator{\Re}{Re}

\renewcommand{\vec}[1]{\boldsymbol{\mathrm{#1}}}

\begin{document}

\title{Peeking under the clouds:\\
Is exoplanet imaging with the solar gravitational lens feasible?}

\titlerunning{Exoplanet imaging under the clouds}

\author{Viktor T. Toth$^1$}

\institute{\vskip 3pt$^1$Ottawa, Ontario K1N 9H5, Canada}


\abstract{
Exoplanet imaging using the solar gravitational lens is an enticing prospect. The fundamental physical properties of the lens, including its angular resolution and light amplification, promise exceptional capabilities. These expectations, however, are tempered by the realization of numerous challenges, including imperfections of the lens itself, noise sources, the properties of the imaging target and difficult technical issues. We discuss, in particular, a subject not previously addressed, the impact of temporally varying surface features, notably a variable cloud cover, obscuring the target exoplanet. This has a substantial detrimental effect on image recovery, leading to our cautious assessment of the practical feasibility of using the Sun's gravitational field as an effective telescope.
}

\maketitle

\section{Introduction}

Compact, massive sources of gravitation such as the Sun deflect electromagnetic waves, including light. Light rays grazing the solar surface are deflected by $\sim$1.75 arc seconds, a prediction of general relativity \citep{Einstein1915b} that was first observationally confirmed by Eddington's famous 1919 solar eclipse expedition \citep{Eddington1920}. Knowing the solar radius, $R_\odot\simeq 6.96\times 10^8$~m allows us to perform a simple calculation: initially parallel rays of light converge in a focal region that is located $\sim$548~AU (astronomical units) from the Sun. This suggests that the Sun, serving as a giant magnifying lens, can be used to view distant sources at levels of light amplification and angular resolution far beyond what is achievable using manufactured instruments.

When we investigate the details, however, a more nuanced picture emerges. The solar gravitational lens (SGL), like any compact gravitational lens, is a lens characterized by significant negative spherical aberration \citep{SGL2017}. It does not project light to a focal point. Rather, light is focused along a focal half-line. Observers at increasing distance from the Sun would see rings of light (the Einstein ring) surrounding the Sun, with the ratio of the observed angular radius of the ring vs. the angular radius of the solar disk increasing as the square root of the observer's distance. As a consequence, the SGL projects a blurred image of a light source even under ideal circumstances \citep{SGL2020c}.

In reality, there are further challenges, which we addressed in past studies. These range from additional distortions due to the small but significant quadrupole mass moment of the Sun \citep{SGL2021a}, various sources of noise and light contamination such as the solar corona \citep{SGL2018e}, the formidable technical difficulties associated with the SGL focal region distance and the need for precise navigation \citep{SGL2021i,SGL2022c}, and the fact that intended targets, exoplanets, will rotate around their axes under varying illumination while being observed \citep{SGL2023d}.

One challenge remained unaddressed: potentially life-bearing Earth-like exoplanets also have a temporally varying appearance, dominated by a variable cloud cover. It is this issue that is the focus of the current study. More broadly, however, it leads to a fundamental question. We know that exoplanet imaging using a gravitational lens is possible in principle. But is it {\em feasible}? That is to say, is it possible to recover images of planets in distant solar systems at a level of quality and resolution that is commensurate with the required effort?

We recap the basic concepts of the SGL in Section~\ref{sec:basics}. To establish context, we discuss in more detail the specific challenges that an SGL imaging mission will have to answer in Section~\ref{sec:imaging}. The cloud cover, the main topic of this paper, is discussed in detail in Section~\ref{sec:clouds}. We assess these results in the context of overall feasibility and offer conclusions in Section~\ref{sec:summary}.

\section{Basic concepts}
\label{sec:basics}

Our study of the SGL began with the simplest model \citep{SGL2018e}: Treating the Sun as a compact, spherically symmetric source of gravitation in the weak field approximation, and in the context of Maxwell's theory of electromagnetism, we developed a model of an electromagnetic wave, passing near, and deflected by, the Sun.

The SGL is characterized by a point-spread function (PSF) \citep{SGL2020c,SGL2020a} that, in the geometric optics limit \citep{SGL2023e}, may be averaged over the circular aperture $d$ of an observing telescope as
\begin{align}
\overline{\tt PSF}(\rho)=\frac{2}{\pi}\Re{\tt E}\left(\frac{\rho}{\tfrac{1}{2}d}\right),\label{eq:PSF}
\end{align}
where ${\tt E}(z)=\int_0^1\sqrt{1-k^2t^2}/\sqrt{1-t^2}~dt$ is the complete elliptic integral in Legendre normal form and $\rho$ is the distance of the center of the telescope aperture, within the SGL's projected image plane (i.e., the telescope lens plane), from the intersection of the  ``optical axis''---the imaginary line connecting the distant source with the center of the Sun and continuing towards the focal region---and the image plane. We derived this expression from a robust wave-optical model of the gravitational lens, but it is also derivable in the context of geometric optics. When $\rho\gg d$, this expression is very closely approximated by $\overline{\tt PSF}(\rho)\simeq d/4\rho$, which makes it explicit that the PSF drops off very slowly, with significant quantities of light spread over the image plane, resulting in substantial blur.

This PSF characterizes lensing by a spherically symmetric mass. The actual Sun has a small quadrupole mass moment and higher multipole mass moments \citep{SGL2021a}, which lead to a modified PSF that incorporates angular dependence: ${\rm PSF}(\vec{x})=|B(\vec{x})|^2$, with the complex amplitude of the electromagnetic field in the lens plane characterized by
\begin{align}
B({\vec x})=\frac{1}{2\pi}\int_0^{2\pi} d\phi_\xi \exp\bigg[&{-i}\frac{2\pi}{\lambda}\Big(\sqrt{\frac{2r_g}{z}}\rho\cos(\phi_\xi-\phi)\nonumber\\
&{}+2r_g\sum_{n=2}^\infty \frac{J_n}{n}\left(\frac{R_\odot }{\sqrt{2r_g z}}\right)^n\sin^n\beta_s\cos[n(\phi_\xi-\phi_s)]\Big)\bigg],
\end{align}
where $\vec{x}=(\rho,\phi)$ is a location in the image plane, $J_n$ is a dimensionless zonal harmonic coefficient characterising the $n$-th mass moment of the Sun, while $\phi_s$ and $\beta_s$ are the heliocentric azimuth and colatitude of the observing location. There is no closed form expression like (\ref{eq:PSF}) in this case involving elementary or well-known special functions, though the corresponding integrals can be evaluated using standard numerical methods. As one might expect, the presence of the quadrupole moment can significantly increase blur at higher imaging resolutions.

The SGL thus projects light from distant sources by forming blurred images of those sources in any image plane that is at a distance greater than $\sim$548~AU from the Sun, perpendicular to the optical axis. The size of the projected image is easily calculated using elementary geometry: given a source distance $R$ and an image plane distance $r$ from the center of the Sun, the size of the projected image will be $r/R$ times the size of the source. By way of example, an exoplanet with the same diameter as the Earth, located at 30 pc from the solar system will have a projected image of $\sim$1,300 meters in diameter at $\sim$650~AU from the Sun.

This geometry dictates any imaging strategy. Rather obviously, we do not have access to a movie theater projection screen that is several square kilometers in size. Instead, we envision a spacecraft, or more likely, several spacecraft forming a semi-autonomous network, each carrying a telescope that is large enough to reliably distinguish the solar disk from the Einstein ring that surrounds it. These spacecraft will then serve as single-pixel instruments, ``light buckets'', sampling the projected image at a large number of locations. Once collected, these samples can be used to reconstruct an image.

While this sounds simple in principle, there are numerous significant challenges, some generic, others more specific to the nature of the most likely imaging targets, Earth-like exoplanets.

\section{Exoplanet imaging with the solar gravitational lens}
\label{sec:imaging}

As we have just seen, the SGL presents us with an amazing promise. The gravitational field of the Sun forms an imperfect magnifying lens with an aperture greater than 1.4 million kilometers, which implies an angular resolution of $10^{-10}$ arc seconds or better at optical wavelengths. The light collecting area of the SGL may be computed as greater than $\pi dR_\odot$, which corresponds to thousands of square kilometers even if an observing telescope of modest aperture, $d=1$~m, is used in conjunction with the SGL. This may lead to the impression that, albeit formidable, only technical challenges (propulsion, navigation, power, communication) remain before the SGL can be put to use to obtain megapixel-resolution images of exoplanets.

\begin{figure}
\begin{center}
\includegraphics{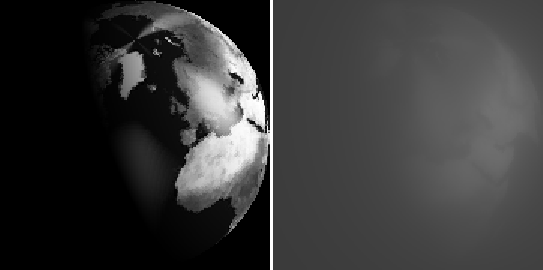}
\end{center}
\caption{\label{fig:projection}The challenge of deconvolution is illustrated by this example. Left: Direct view of an Earth-twin exoplanet corresponding to a vantage point above 45~N latitude, with the planet partially illuminated by its host star and lightly obscured by simulated clouds. Right: The blurred projection of the same by the SGL. Note that the image on the right has been brightened for visibility, otherwise it would appear in print as a nearly black rectangle. Its contrast has not been manipulated, however.}
\end{figure}

Alas, this is not the case. Apart from the technical challenges, which are indeed formidable, there turned out to be some fundamental limitations of the SGL itself, which we discussed at length in past studies. These include:
\begin{itemize}
\item Even a spherically symmetric gravitational lens is an imperfect lens with negative spherical aberration. The lens projects a blurred image into its projected image plane (Figure~\ref{fig:projection}). Removing the blur, i.e., deconvolving the image, incurs a substantial noise penalty.
\item The Sun, though almost perfectly spherical, has a nonvanishing quadrupole moment. The dimensionless magnitude of this moment is of ${\cal O}(10^{-7})$, which does not appear significant until we consider that this factor multiplies the amount by which light, grazing the solar disk, is deflected to reach the SGL focal line in the focal region. This amounts to a deflection greater than $R_\odot$, and consequently, even a quadrupole moment as small as  ${\cal O}(10^{-7})$ may displace light by several hundred meters in the projected image plane. Unscrambling the resulting blur is difficult and significantly increases the amplification of noise by the deconvolution process.
\item The light that arrives from an exoplanet forms an Einstein ring around the Sun as viewed from the SGL focal region. This Einstein ring is situated within the solar corona, which is much brighter. Even if corona light can be removed accurately (e.g., by observing the corona from a second spacecraft some distance away from the first), shot noise remains, which can only be mitigated by very long observational times. The contrast ratio of the Einstein ring vs. the corona also requires optical sensors of possibly unrealistic sensitivity.
\item The Einstein ring that is seen may itself be contaminated by light from the exoplanet's host star and possibly, other background sources (distant stars).
\item The location of the exoplanet's projected image in the image plane moves non-inertially, as the optical axis, connecting the exoplanet and the center of the Sun, pivots at the Sun. Especially while observing using longer observation times, it is essential to follow this noninertial trajectory to ensure that the observing instrument receives light from the same region of the exoplanet.
\item The observing spacecraft must be able to ascertain its position in the projected image plane relative to the exoplanet with meter-scale precision or better, despite the lack of any local reference points.
\item The exoplanet does not stand still. It may be rotating like the Earth, and orbiting its host star. This not only limits observational times before motion-induced blur sets in making deconvolution challenging or impossible, but also reduces the illuminated area of the exoplanet that is visible from the solar system at any given time.
\item The exoplanet's appearance may change. The visual appearance of the Earth is dominated by its rapidly changing cloud cover; we expect a potentially life bearing exoplanet to be no different in this regard.
\end{itemize}

All of these challenges have been addressed in our past studies, except the last one: the variable appearance of the exoplanet due to its time-varying surface or atmospheric features.

In particular, we have learned how to deal with the predictable temporal changes in an exoplanet's appearance due to planetary rotation and variable illumination during its orbit around its host star \citep{SGL2023d}. To recap, we have developed a method that combines the blur due to the SGL with the temporally varying appearance of the source. Thinking of the planet + SGL system as a single instrument that maps the entire surface of the exoplanet onto the SGL projected image plane, we arrived at an effective point-spread function, or PSF. As this PSF depends not just on the position within the projected image plane but also on time, deconvolution is no longer possible using computationally efficient Fourier-optics methods. Instead, we did it the hard way: After modeling the combined effects of the SGL, planetary rotation and illumination, we used standard computational methods to invert the process and solve numerically for the source, which is to say, the topography of the exoplanet.

This approach proved fruitful albeit somewhat discouraging. Using the Earth as a stand-in, we were able to recover its surface topography, but the recovered image had substantial noise, with a signal-to-noise ratio ${\tt SNR}\lesssim 2$.

Without a doubt, this figure can be improved by more innovative data acquisition and image reconstruction strategies. These may include, e.g., restricting the sampling to regions of the projected image of the exoplanet that correspond to fully illuminated areas, varying the sampling density by latitude, perhaps even introducing variable weights assigned to the samples.

While these strategies may be worth exploring in future studies, our current goal was to explore the major remaining issue: variable surface features, specifically variable cloud cover.

\section{Dealing with clouds}
\label{sec:clouds}

Imagine a spacecraft that has all it takes: a powerful telescope with a Lyot coronagraph, power, communication, maneuverability, even the ability to maintain precise position within the projected image of an exoplanet in the SGL focal region. It has a simple task: Moving from position to position in the image plane, spend some time at each pixel location, collecting enough light to be able to reliably distinguish signal (the brightness of a faint Einstein ring) from noise (mostly light from the solar corona).

In a perfect world, this would be enough: Once our spacecraft collects a sufficient number of samples, our work is done. Applying a deconvolution algorithm, we will have obtained a good quality resolved image of the target.

Unfortunately the world is not perfect. Apart from all the issues that we addressed previously, the exoplanet's surface itself is partially obscured by a rapidly varying cover of clouds. How can we deal with this?

\begin{figure}
\begin{center}
\scalebox{-1}[1]{\includegraphics[scale=1.5]{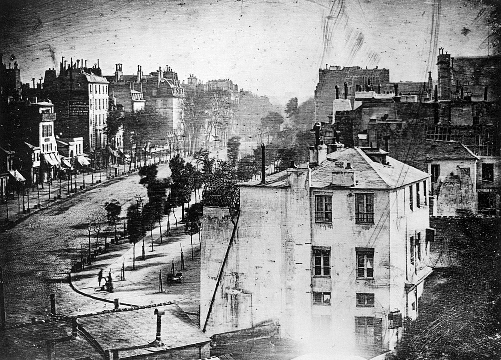}}
\end{center}
\caption{\label{fig:daguerre}The earliest dated daguerreotype showing recognizable humans ({\em Boulevard du Temple} by Louis Daguerre, from 1837--38 \citep{daguerre1838boulevard}). The streets and sidewalks appear empty. In reality, they were probably bustling with traffic, but long periods of exposure erased the presence of any moving person, animal or vehicle as mere random noise. Only the stationary figures of a passer-by getting a shoe shined and the person tending to the shoe remain.}
\end{figure}

Surprisingly, we find a possible solution by looking at one of the earliest photographs in existence: a daguerreotype created by Louis Daguerre, taken in 1837 or 1838 \citep{daguerre1838boulevard}, showing a section of a Paris city street from his laboratory window (Figure~\ref{fig:daguerre}). This image is often described as the first ever photographic image showing human beings. They are not easy to spot, however: they are tiny stick figures on a sidewalk, judging by their posture, likely a person having a shoe shined by a shoe shining street vendor. Yet what is even more remarkable is what is not seen in this image. It is broad daylight. Probably an ordinary weekday. There are businesses open. The wide street is open for traffic. Yet not a soul is seen. No passers-by, no horses or other animals, no carriages, no people. Why?

The most likely answer is also the simplest: They have been averaged out by the multi-minute exposure time necessitated by Daguerre's primitive photographic technology. (Something similar is also seen often in 20th century long-exposure night photographs of city scenes, in which the headlights of vehicles are seen as long streaks of light, but the vehicles themselves appear translucent or entirely invisible.)

Going back to our spacecraft observing the exoplanet, an observational campaign will likely require a long time. The spacecraft will linger at each pixel for minutes before moving on to the next; collecting data from tens of thousands of pixels or more will likely take days, weeks, perhaps months or more. During this time, the clouds move, dissipate, and new clouds are created. The cloud cover changes often and rapidly. Averaged over time, it turns into random noise, much like the passers-by in Daguerre's iconic photograph.

This, then, is our expectation: an SGL campaign will be able to ``peek under the clouds'', the cloud cover contributing nothing other than random noise.

Of course random noise is still unwelcome news, especially when we are already dealing with abysmal SNRs, after having taken into account issues like noise from the solar corona or blur due to the solar quadrupole moment or the planet's rotation and variable illumination. Still, we have to play with the cards Nature dealt us. Let us investigate if it is still possible, at least in principle, to recover useful images with the SGL.

We begin by characterizing the cloud cover at any point in the source plane using the opacity $\tau$ of the clouds, a dimensionless number between $\tau = 0$ (no clouds at all) and $\tau = 1$ (fully covered by clouds forming a 100\% reflective Lambertian surface). The signal (light from the surface) is reduced by the factor $1-\tau$, whereas noise with an amplitude proportional to $\tau$ is added to this signal. As this noise is strictly additive, it is equivalent to introducing a bias of $\tfrac{1}{2}\tau$ and a noise amplitude of $\pm\tfrac{1}{2}\tau$. Treating as the ``signal'' the exoplanet image with all other noise sources included, the cloud cover therefore contributes an additional SNR of $(1-\tau)/\tfrac{1}{2}\tau$.

In previous studies \citep{SGL2020c,SGL2020f}, we arrived at a useful expression for the deconvolution penalty associated with a spherically symmetric lens:
\begin{align}
\frac{{\tt SNR}_{\tt R}}{{\tt SNR}_{\tt C}}\simeq 0.891\frac{D}{d\sqrt{N}},
\end{align}
where ${\tt SNR}_{\tt R}$ is the recovered signal while ${\tt SNR}_{\tt C}$ characterizes the noise present in the blurred projection before deconvolution; further, $d$ represents the observing telescope's aperture, $D$ is the spacing between pixel observations (if the image plane is fully covered by adjacent pixels, $D=d$) and $N$ is the total number of observed pixels.

We may also think of this penalty as a term that amplifies noise that is present before deconvolution. We therefore expect this deconvolution penalty to also apply to the SNR characterizing the average cloud cover, resulting in a post-deconvolution SNR contribution due to clouds in the form
\begin{align}
{\tt SNR}_{\tt cloud}(\tau)\simeq  0.891\frac{D}{d\sqrt{N}}\frac{1-\tau}{\tfrac{1}{2}\tau}.
\end{align}

To check the approximate validity of this estimate and indeed, to verify that it is, in fact, possible to recover surface topography beneath a cloud cover using data from SGL-type observations made over an extended time period, we modified software \citep{TEMPORAL2024}, previously used to model SGL imaging of an orbiting, spinning exoplanet \citep{SGL2023d}, by introducing a simulated cloud cover. Our goal was not to model the climate, of course, just to introduce a simple cloud model with the anticipated statistical properties. To this end we constructed a heuristic multilayer cloud model using a Perlin noise algorithm \citep{perlin1985}, with clouds forming, dissipating, and moving with a linear velocity across the cylindrical map projection. We validated the model by ascertaining that the average cloud cover remained stable over numerous long simulation runs and that over time, all regions of the topography were covered by clouds at an equal rate.

With the cloud model in place and integrated with our SGL image recovery simulation, we were ready for the next step. Before we could proceed and make sense of any simulation results, however, we need to pause for a moment. In this simulation, there is no ``telescope''. We sample the projected image plane without modeling the sampling instrument. What would our telescope aperture $d$ be like? What value can we use in place of $D/d$? Our question is answered by the fact that we used a relatively low resolution data source, a cylindrical projection of the Earth's topography at a resolution of 1 degree (latitude and longitude). This is our ``source pixel size'', the closest approximation of the telescope aperture $d$. Our vantage point corresponding to the latitude 45$^\circ$~N, pixels below 45$^\circ$~S are excluded from the process. Thus we downsample $360\times 135$ source pixels to a reconstructed image that, in our case, contains $\sim$5,700 pixels. At the same time, we must also recall that there was a factor of $\pi/4$ in the original deconvolution penalty estimate \citep{SGL2020c}, corresponding to a circular image in a rectangular imaging region. This is no longer needed, as our cylindrical projection is itself rectangular. As a result, we adopt the value $D/d=(4/\pi)\sqrt{360\times 135/5700}\simeq 3.31$ in our deconvolution penalty estimate for cloud noise, which yields
\begin{align}
{\tt SNR}_{\tt cloud}(\tau)\simeq 0.0439\left(\sqrt{\frac{n}{48600}}\right)\left(\frac{5700}{N}\right)\frac{1-\tau}{\tfrac{1}{2}\tau},
\end{align}
where $n$ is the number of source pixels.

\begin{figure*}
\begin{center}
\includegraphics[scale=2.25]{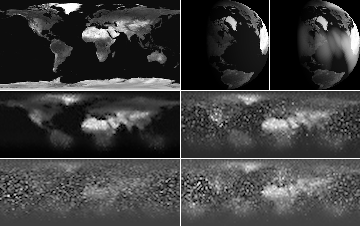}
\end{center}
\caption{\label{fig:example}Reconstructed topography of an orbiting, spinning Earth-twin exoplanet using $\sim$24,000 simulated measurements of its SGL-projected image. The simulation entailed 200 measurements taken at 2.5 hour intervals obtaining 120 data points each, over a full duration of 500 hours. As the vantage point is directly above the $45^\circ$~N latitude, the topography below $45^\circ$~S is permanently obscured and thus excluded from the reconstruction. The reconstruction produced a half-resolution image of $\sim$5,700 pixels. Top left: The original topography. Top right: Snapshots of the same, with illumination corresponding to Julian date 2460087, with no clouds and with simulated $\sim$13.7\% cloud cover. Bottom two rows, clockwise from upper left: 1) the topography, reconstructed from a simulation with no clouds; 2) The same, with a $\sim$2.7\% cloud cover; 3) $\sim$6.8\% cloud cover; and 4) $\sim$13.7\% cloud cover. Images generated using \citep{TEMPORAL2024}.}
\end{figure*}

Without further ado, let us peek at simulation results, such as those shown in Figure~\ref{fig:example}. Our goal is to reconstruct the top image (topographic view of the exoplanet) from numerous observations like those shown in Fig.~\ref{fig:projection}, with the planet rotating, orbiting its host star, and featuring a time-varying cloud cover. We ran four cases, including a baseline case with no cloud cover, gradually increasing the cloud cover from 2.7\% through 6.8\% to 13.7\% (these values corresponded to incremental values of an internal variable that controlled the cloud cover simulation.) The simulation covered a period of 500 hours, sampled at 2.5 hour intervals, with $\sim$120 samples each. The samples were placed in a regular grid but with random small displacements, covering the entire rectangular projected image area, which of course implies some ``wasted'' samples, centered on locations where the planet was not illuminated.

Without cloud cover, our simulation yielded ${\tt SNR}\simeq 1.86$ in the run corresponding to the upper left image in the bottom section of Figure~\ref{fig:example}. After we turned on clouds, the value dropped, as shown in Table~\ref{tb:penalty}.
This SNR drop is broadly consistent with our estimate, treating the noise present in the cloud-free case and the noise due to cloud cover as uncorrelated values:
\begin{align}
{\tt SNR}(\tau)=\big[{\tt SNR}_{\tt cloud}(\tau)^{-2} + {\tt SNR}_{\tt clear}^{-2}\big]^{-1/2}.
\end{align}
As we can see from Table~\ref{tb:penalty}, our expression overestimates the SNR for very light cloud cover, but as the cloud cover nears 10\%, it turns into an underestimate. This may be due to the fact that SNR estimates of 1 or less are generally unreliable, or perhaps that treating noise sources as uncorrelated and Gaussian is not wholly justified. Nonetheless, the broad trend is unambiguous: increasing cloud cover substantially reduces the SNR of an already noisy recovered image of the planetary surface, making it very difficult to obtain useful reconstructions of the topography.

\begin{table}
\caption{\label{tb:penalty}Computed SNR values and values obtained from simulation using increasing cloud cover ($\tau$).}
\begin{center}
\begin{tabular}{c|c|c}
~&\multicolumn{2}{c}{SNR}\\
$\tau$&computed&simulated\\\hline\hline
~0.0\%&~&1.86\\
~2.7\%&1.60&1.44\\
~6.8\%&1.01&1.13\\
13.7\%&0.53&0.80\\\hline
\end{tabular}
\end{center}
\end{table}

\section{Are alternative strategies possible?}

The preceding results are predicated on a specific imaging strategy: using a family of observing telescopes as ``light buckets'', scanning a multi-kilometer image plane one ``pixel'' at a time, collecting light and then moving on, performing a slow scan of the projected image.

Thinking about the geometry of the problem, one may wonder: What if the projected image area was smaller, perhaps as a result of choosing a more distant target? Could it be fitted within the field of view of a single instrument? And must we use the telescope as a single-pixel light bucket? Could we not make use of additional information that might be present in a resolved Einstein ring observed around the Sun, perhaps by segmenting the ring to the extent permitted by the resolution of the observing telescope? What about spectroscopy?

The answer is no to these questions, for reasons that we addressed in prior studies, but which may be worth recapping here: corona noise \citep{SGL2018e}, the navigational problem \citep{SGL2021i}, and last but not least, the solar quadrupole moment \citep{SGL2021a}.

Corona noise represents a most serious challenge. The solar corona is several orders of magnitude brighter than the faint Einstein ring due to light from an exoplanet. Even if the corona can be accurately measured and removed (e.g., by observing it from `off-axis' spacecraft that see only the corona, not the exoplanetary Einstein ring), it remains an inevitable contributor to shot noise. Any attempt to image a fainter target (such as a more distant exoplanet), segmenting the Einstein ring, or splitting the observed signal by wavelength would amplify this noise contribution, which is proportional to the square root of the number of photons collected per sample.

At the root of the navigational problem is the fact that the image projected by the SGL into the imaginary image plane moves noninertially. This motion represents the combined effect of the exoplanet's orbit around the planet-host star two-body system barycenter, perturbed by other bodies in the same system, and then projected by the Sun, which moves relative to the exoplanet host star, with its motion itself perturbed by the gravitational pull of planets within our own solar system. In our envisioned scenario, observing spacecraft would enter the multi-kilometer projected image region in the image plane, performing prescribed maneuvers with precision navigation relative to a local inertial reference frame provided by other spacecraft in the constellation. This would allow for a later reconstruction of the observing spacecraft trajectories within the image and thus, image deconvolution. This is not possible if the projected image is very compact. In this case, the observing instrument would have to find the precise location of this projection, which is to say, perform sub-meter scale precision navigation with no nearby reference. Such navigation is vastly beyond any present or foreseeable navigational capabilities in deep space in the outer solar system.

Last but not least: Such a compact image does not even exist. The Sun, unfortunately, is not a perfect gravitational monopole. Its quadrupole moment is small in relative terms, but over the span of 650 AU, even that small quadrupole contribution is sufficient to deflect light by up to several hundred meters. In the case of a large, multi-kilometer image of a nearby exoplanet, this additional blur makes deconvolution harder, but not impossible. A more compact image, however, would be washed out altogether by the quadrupole blur and would be severely affected even by higher-order mass multipoles of the Sun.

\section{Discussion}
\label{sec:summary}

It has been in light of these results, in particular the bottom left image of Figure~\ref{fig:example}, that the question in the title of our study becomes pertinent: Is gravitational lensing with the SGL feasible?

As the images show, vague but recognizable outlines of the major continents become visible despite the noise. Granted, this simulation used only light cloud cover. In contrast, the actual cloud cover of the Earth is close to 70\%, substantially greater than the model we used. With such an extensive cloud cover, our simulation could not recover a meaningful image of the planetary topography.

Improvements are possible, as we discussed. These may improve optimized sampling strategies, a better reconstruction algorithm, filtering of noise, incorporation of climate models that constrain the nature and shape of cloud cover, perhaps even machine learning methods.

On the other hand, there are several practical challenges that will prove formidable.
These include simply finding the SGL focal location that corresponds to a specific, faint target, the required ultra-precise navigation to stay on target, dealing with light contamination from other sources, dealing with imaging a faint Einstein ring over a bright solar corona, and dealing with an even worse deconvolution penalty affecting the signal-to-noise ratio due to the quadrupole mass moment of the Sun.

Perhaps these challenges can be overcome. Which leads us to the next question: Is it worth it? With present-day technology, an SGL mission is a multidecadal endeavor. Pushing the limits of available technology, e.g., solar sailing with very close perihelia and in-flight assembly of a large telescope from microspacecraft accelerated by large solar sails, it will still require 25 years or more to reach the SGL focal region \citep{SGL2022c,SGL2023c}. At first sight, the promise of the SGL was enticing: megapixel-resolution images, revealing surface features of 10~km or less, allowing us to see not just topography but weather, vegetation, oceans and ice sheets, perhaps even signs of civilization. Instead, we are left with images such as those shown at the bottom of Figure~\ref{fig:example}: grainy, low-resolution black-and-white projections with the faint outlines of continents only barely visible.

Can such a result compete against ground-based observations or observations conducted using Earth-orbiting spacecraft? There are other proposed methods of exoplanet imaging. Conceivably, a combination of time-sequence observations from multiple instruments might lead to reconstructed images of exoplanet surface features that are comparable to what we seem to be able to achieve using the SGL, but without the risks and technical complications associated with an SGL mission.

Are there alternative imaging strategies possible, perhaps exploring features of the SGL that we have not considered, such as recovering details from the Einstein ring itself? These proposals run into fundamental limitations with respect to solar corona noise, navigation and the additional blur due to the solar quadrupole moment.

And yet, these results perhaps point in a different direction. Perhaps, responding to how we posed the question in the Introduction, we might wish to look beyond the feasible and investigate what might be possible. Many of the problems we encountered---navigation, stationkeeping, reconstruction of topographic features of a spinning, orbiting exoplanet with variable surface features---arise because of the fundamental limitations imposed by the proposed use of at most a small number of spacecraft to physically scan, pixel-by-pixel, a projected image plane area that may be many square kilometers in size. Yet even today, we already have satellite networks featuring thousands of spacecraft (as of the time of this writing, the Starlink network has over 6,000 operational vehicles in orbit.) Granted, these are Earth-orbiting satellites under ground control, utilizing solar power as their electric power source, not a satellite network 650~AU from the Sun, operating autonomously using the only electric power source that is suitable for such a mission, nuclear power. Yet it seems that at least in principle, launching and operating such a large network of satellites, even in deep space in the far reaches of the outer solar system, is possible. A network of 10,000 satellites, each equipped with a meter-class telescope, can ``snap'' an instant photograph of an exoplanet by simultaneously sampling the projected image plane at $100\times 100$ grid locations. Many of the issues with motion, tracking, stationkeeping, variations in surface features and illumination, vanish. Some issues remain: the signal-to-noise ratio is still abysmal due to the solar corona, and the image blurred by an imperfect gravitational lens still requires deconvolution. But these problems are much easier to address if the entire projected image can be sampled at the same time. On the other hand, such a satellite network can also utilize spectrography, not to mention that it can take repeated images, thus obtaining motion pictures of the target exo-world.

For now, a 30-40 year mission to a heliocentric distance of 650~AU or beyond will likely remain in the realm of fiction, especially if it involves a very large network of satellites. However, if we discovered an exoplanet with unmistakable signs of industrial civilization, it is easy to see how such an enormous deep space mission might gain broad international support. It might, after all, represent the first time in our history that we can confirm the existence of a civilization other than ours.


\section*{Acknowledgments}

VTT acknowledges the generous support of Vladimir Andreev, Plamen Vasilev and other Patreon patrons.

\bibliography{refs}

\end{document}